\documentclass[fleqn,usenatbib]{mnras}
\usepackage{newtxtext,newtxmath, xcolor}
\usepackage[T1]{fontenc}
\DeclareRobustCommand{\VAN}[3]{#2}
\let\VANthebibliography\thebibliography
\def\thebibliography{\DeclareRobustCommand{\VAN}[3]{##3}\VANthebibliography}

\usepackage{graphicx}	
\usepackage{amsmath}	
\usepackage{multirow}
\usepackage{hyperref}
\usepackage{appendix}

\newcommand{\NumAGNField}[1]{203}
\newcommand{\OverlapAGN}[1]{42}

\title[The AGN Optical Variability Fundamental Plane]{The AGN Optical Variability Fundamental Plane}

\author[A. P. Tarrant et al.]
{Ashley P. Tarrant,$^{1,2}$\thanks{E-mail: tarrant.11@buckeyemail.osu.edu}
Jason T. Hinkle,$^{3,4}$
Benjamin J. Shappee,$^{3}$
Christopher S. Kochanek,$^{1,5}$
\newauthor
Daniel R. Hey,$^{3}$
Connor Auge,$^{6}$
Anna V. Payne,$^{7}$
Michael J. Bolish,$^{3}$
Heechan Yuk,$^{8}$
Xinyu Dai,$^{8}$
\newauthor
Katie Auchettl,$^{9,10}$ 
Todd A. Thompson,$^{1,5}$ and 
Helena Treiber$^{11}$ 
\\
$^{1}$Department of Astronomy, Ohio State University, 140 W. 18th Ave, Columbus, OH 43210\\
$^{2}$Department of Physics, Ohio State University, 191 W. Woodruff Ave, Columbus, OH 43210\\
$^{3}$Institute for Astronomy, University of Hawai'i, 2680 Woodlawn Dr, Honolulu, HI 96822\\
$^{4}$NASA FINESST FI \\
$^{5}$Center for Cosmology and Astro-Particle Physics, Ohio State University, 140 W. 18th Ave, Columbus, OH 43210\\
$^{6}$ Eureka Scientific, 2452 Delmer Street, Suite 100, Oakland, CA 94602-3017 \\
$^{7}$Space Telescope Science Institute, 3700 San Martin Drive, Baltimore, MD 21218 \\
$^{8}$Homer L. Dodge Department of Physics and Astronomy, University of Oklahoma, 440 W. Brooks St., Norman, OK 73019\\
$^{9}$School of Physics, The University of Melbourne, Parkville, VIC 3010, Australia\\
$^{10}$Department of Astronomy and Astrophysics, University of California, Santa Cruz, CA,  95064, USA\\
$^{11}$Department of Astrophysical Sciences, Princeton University, 4 Ivy Lane, Princeton, NJ 08544
}

\date{Accepted XXX. Received YYY; in original form ZZZ}

\pubyear{2025}

\begin{document}

\label{firstpage}
\pagerange{\pageref{firstpage}--\pageref{lastpage}}
\maketitle

\begin{abstract}

We investigate the relationship between AGN optical variability timescales, amplitudes, and supermassive black hole (SMBH) masses using decade-long homogeneous light curves from the All-Sky Automated Survey for SuperNovae (ASAS-SN). We fit a damped random walk (DRW) model to high-cadence, long-baseline ASAS-SN light curves to estimate the characteristic variability timescale ($\tau_\text{DRW}$) and amplitude ($\sigma$) for 57 AGN with precise SMBH mass measurements from reverberation mapping and dynamical methods. We confirm a significant correlation between $\tau_\text{DRW}$ and SMBH mass, and find: $\textrm{log}_{10}(M_{\textrm{BH}}/ \textrm{M}_{\odot}) = (1.85 \pm 0.20) \times \textrm{log}_{10} ({\tau \textsubscript{DRW} / 200 \textrm{ days}}) + 7.59 \pm 0.08$. Incorporating $\hat{\sigma}^2 = 2\sigma^2/\tau_{\textrm{DRW}}$ in a plane model fit yields significantly improved residuals, and we find: $\textrm{log}_{10}(M_\text{BH}/ \textrm{M}_{\odot}) = (2.27 \pm 0.20)\times \textrm{log}_{10} (\tau_\text{DRW} / 200 \textrm{ days}) + (1.20 \pm 0.20)\times \textrm{log}_{10} (\hat{\sigma} / \textrm{1 mJy/days}^{1/2}) + 7.68 \pm 0.08$ with a typical scatter of 0.39 dex. We calculate $\tau_\text{DRW}$, $\hat{\sigma}$, and estimate SMBH masses for \NumAGNField{} bright ($V<16$ mag) AGN from the Milliquas catalog and compare these estimates with existing measurements from the BAT AGN Spectroscopic Survey for \OverlapAGN{} overlapping AGN for validation. Assuming this relation, future surveys like the Legacy Survey of Space and Time (LSST) will extend this method to survey SMBHs in the range $7\lesssim\textrm{log}_{10}({M_\text{BH}/M_\odot})\lesssim 9$ out to $z\sim1$ and $\textrm{log}_{10}({M_\text{BH}/M_\odot})\sim8.0$ out to $z\sim4$. By the end of the 10-year LSST, ASAS-SN could have 25 years of data probing SMBHs from $5\lesssim \textrm{log}_{10}({M_\text{BH}/M_\odot})\lesssim 10.5$ in the local universe and $\textrm{log}_{10}({M_\text{BH}/M_\odot})\sim9.0$ out to $z\sim2$. Measuring AGN variability with these datasets will provide a unique probe of SMBH evolution by making estimates of $M_\text{BH}$ spanning several orders of magnitude with photometric observations alone.

\end{abstract}




\section{Introduction}
Supermassive black holes (SMBHs) exist in the centers of all massive galaxies \citep[e.g.,][]{Rees1998,KormendyRichstone1995,Magorrian1998}, and there are several observed correlations between SMBH and galaxy properties \citep[e.g.,][]{KormendyHo2013} that suggest that SMBHs and their hosts evolve together. Some of these black holes have an accretion disk of infalling gas, creating an active galactic nucleus \citep[AGN; e.g.,][]{Antonucci1985,Urry1995,Netzer2015}. Measuring the mass of the central SMBH is critical to understanding the structure of AGN and accretion disk physics, since most properties of the accretion disk scale with this mass. 

However, SMBH masses ($M_{\textrm{BH}}$) are difficult to measure. Recently two SMBHs, Pōwehi in M87 and Sgr A$^*$ in the Milky Way, have been exquisitely measured by direct imaging \citep{EHT2019, EHT2022SgrA}, but the required angular resolutions limit this method. For nearby (10s of Mpc) galaxies, it is possible to measure the kinematics of stars and gas inside the SMBH's sphere of influence \cite[e.g.,][]{Barth2001,McConnell2012} and constrain $M_{\textrm{BH}}$. Masses of active SMBHs can be measured with a third method, reverberation mapping (RM), which trades the spatial resolution requirements of the first two methods for temporal resolution by using the light travel delay between the luminous accretion disk and the broad line region to estimate the distance. This distance is then combined with broad emission line widths as a proxy for orbital velocity, leading to an estimate of $M_{\textrm{BH}}$ \citep[e.g.,][]{Blandford1982,Peterson2004,Onken2004,Vestergaard2006,Bentz2013}. AGN RM measurements require long observational campaigns and expensive spectroscopic observations \citep{Horne2004}. These methods for measuring $M_{\textrm{BH}}$ have allowed us to calibrate the relationships between the $M_{\textrm{BH}}$ and host-galaxy properties such as the M-$\sigma_{*}$ relation \citep[e.g.,][]{FerrareseMerritt2000,Gebhardt2000,Onken2004,Gultekin2009} and single-epoch spectral mass measurements \citep[e.g.,][]{Kapsi2000,Kapsi2005,Bentz2009a,Bentz2013}. 

Here we aim to calibrate a relatively inexpensive and widely-applicable relation for estimating the masses of SMBHs in AGN. It is well known that AGN vary stochastically across the electromagnetic spectrum \citep[e.g.,][]{Ulrich1997, vandenBerk2004}. This variability occurs on both short (less than a day) and long (many months) timescales. The power spectral density (PSD) of AGN is often well-characterized as a broken power-law $P(f)=f^{-\alpha}$ with a break between red noise at high frequencies with $\alpha \sim2$ \citep[e.g.,][]{Giveon1999,Collier2001} and white noise at low frequencies with $\alpha \sim0$ \citep[e.g.,][]{Kozlowski2010,MacLeod2010}. The driving mechanism behind these fluctuations is not well understood, though it is thought to be related to thermal variations in the accretion disk \citep[e.g.,][]{Siemiginowska1989,Kawaguchi1998,Collier2001,Kelly2009}. Recently, multi-band light curves mapped to temperature fluctuations on a disk revealed that these fluctuations exist as both in-moving and out-moving waves and cannot be fully explained by a simple lamppost model \citep[e.g.,][]{Neustadt2022,Neustadt2024}. Previous works \citep[e.g.,][]{Collier2001,Kelly2009,MacLeod2010,Burke2020, Yuk22} have shown that continuum variability parameters correlate with both the mass and luminosity of the SMBH. Thus, measuring optical variability may also be an efficient and effective way to estimate $M_{\textrm{BH}}$ for AGN \citep[e.g.,][]{MacLeod2010,Burke2020}. 

The variability of AGN is stochastic in nature and is reasonably well-described by the damped random walk (DRW) continuous auto-regressive moving average (CARMA) process \citep{Kozlowski2010,MacLeod2010,MacLeod2012,Zu2011,Zu2013,Kelly2014,Zu2016}. Instead of computing and fitting a PSD, which is sensitive to noise and biased by uneven time sampling, a CARMA process can be fit directly to an irregularly sampled light curve including measurement errors in its likelihood estimation. One popular implementation is JAVELIN \citep{Zu2011,Zu2013}, a package that models the time lag of single or multiple spectroscopic emission lines assuming a DRW process for interpolation. 

\begin{table*}
  \centering
  \caption{Data from the AGN BH Mass Database detailing our initial calibration sample. The first 10 sources are shown here to demonstrate its form and content. The table is included in its entirety as an ancillary file.}
  \label{table1}
  \renewcommand{\arraystretch}{1.5}
  \begin{tabular}{|c|c|c|c|c|c|}
  
    \hline
    \textbf{Object} & \textbf{Right Ascension} & \textbf{Declination} & \textbf{Redshift} & \textbf{log$_{10}(M \textsubscript{BH}/\textrm{M}_{\odot})$} & \textbf{Mass Reference} \\
    \hline
    Mrk 335 & 00:06:19.5 & +20:12:10 & 0.026 & 7.23$^{+0.04}_{-0.04}$ & \cite{Grier2017} \\
    Mrk 1501 & 00:10:31.0 & +10:58:30 & 0.089 & 8.07$^{+0.12}_{-0.17}$ & \cite{Grier2017} \\
    PG0026+129 & 00:29:13.6 & +13:16:03 & 0.142 & 8.49$^{+0.10}_{-0.12}$ & \cite{Kapsi2000,Peterson2004} \\
    PG0052+251 & 00:54:52.1 & +25:25:38 & 0.154 & 8.46$^{+0.08}_{-0.09}$ & \cite{Kapsi2000,Peterson2004} \\
    Fairall 9 & 01:23:45.8 & $-$58:48:21 & 0.047 & 8.30$^{+0.08}_{-0.12}$ & \cite{Peterson2004} \\
    Mrk 590 & 02:14:33.5 & $-$00:46:00 & 0.026 & 7.57$^{+0.06}_{-0.07}$ & \cite{Peterson1998a,Peterson2004} \\
    3C120 & 04:33:11.1 & +05:21:16 & 0.033 & 7.74$^{+0.04}_{-0.04}$ & \cite{Grier2017} \\
    H0507+164 & 05:10:45.5 & +16:29:56 & 0.018 & 6.88$^{+0.07}_{-0.42}$ & \cite{Stalin2011} \\
    Ark 120 & 05:16:11.4 & $-$00:08:59 & 0.033 & 8.07$^{+0.05}_{-0.06}$ & \cite{Peterson1998a,Peterson2004} \\
    MCG+08-11-011 & 05:54:53.6 & +46:26:22 & 0.020 & 7.29$^{+0.05}_{-0.05}$ & \cite{Fausnaugh2017} \\
    \hline
  \end{tabular}
\end{table*}

\begin{figure*}
\centering
    \includegraphics[width=\textwidth]{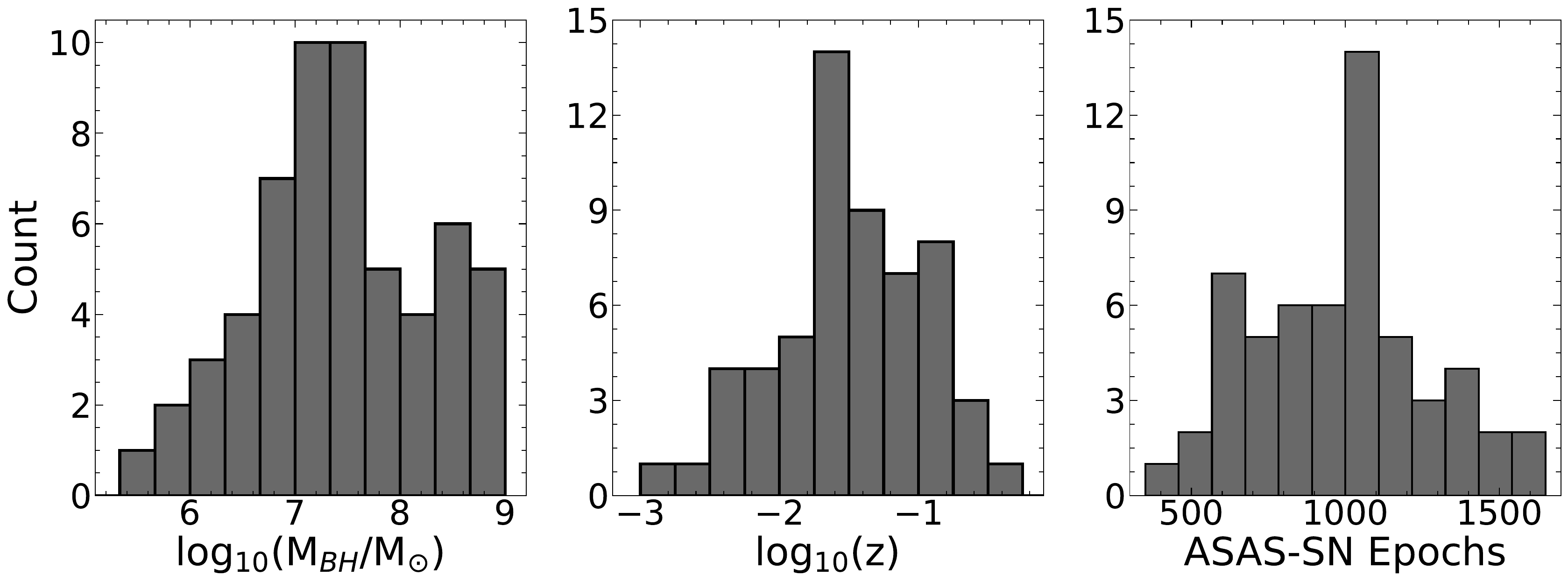}
    \caption{Distributions of the 57 AGN that make up our final calibration sample in mass (left), redshift (center), and number of ASAS-SN epochs (right). The calibration sample consists mostly of nearby Seyfert 1 galaxies. The BH masses span more than three orders of magnitude and the median number of light curve epochs is 1,013.}
    \label{fig:sample_distr}
\end{figure*}

Among the difficulties in measuring the characteristic variability timescale $\tau \textsubscript{DRW}$ is the need to have a long temporal baseline to minimize biases that tend to lead to underestimates of $\tau \textsubscript{DRW}$ \citep{Kozlowski2021,Suberlak2021}. Many previous studies did not use light curves with a sufficient temporal baseline \citep[e.g.,][hereafter KBS09]{Kelly2009} or spliced together light curves from multiple data sets \citep[][hereafter B21]{Suberlak2021,Burke2021}. All-sky surveys such as the All-Sky Automated Survey for Supernovae \citep[ASAS-SN;][]{Shappee2014,Kochanek2017,Hart2023} address both of these concerns, having monitored the entire night sky for over a decade with a near-daily cadence, to provide excellent data for characterizing the variability of a large sample of sufficiently bright AGN. Here we present the first homogeneous sample of long-baseline AGN light curves used to estimate $\tau \textsubscript{DRW}$ and $\sigma$ for objects with well-measured masses. By reducing systematic biases in our data selection, we are able to robustly calibrate the relationship between AGN variability and SMBH mass.

This paper is organized as follows. In Section \ref{data}, we present our calibration sample selection and the observations from the ASAS-SN survey. In Section \ref{est_tau}, we describe our methods for estimating the characteristic timescale and amplitude using the DRW process and present the variability\textendash mass plane. We then calculate masses for \NumAGNField{} bright field AGN using ASAS-SN photometry in Section \ref{calc_bh_mass}. We discuss the future utility of our relation in Section \ref{discussion}. Finally, we summarize this paper in Section \ref{conclusion}. 


\section{Data}
\label{data}
To minimize systematic biases in the variability\textemdash $M\textsubscript{BH}$ relationship, we only use systems with $M_{\textrm{BH}}$ measurements from RM campaigns or spatially resolved gas or stellar kinematics as described in Section \ref{sample}. In Section \ref{observations}, we describe the ASAS-SN survey and how we obtain the light curves used in this study. In Section \ref{measure_var} we investigate ASAS-SN's ability to detect intrinsic variability in these sources and in Section \ref{err_corr} we discuss our handling of the uncertainties in the ASAS-SN data. 

\subsection{Calibration Sample and M$_{\textrm{BH}}$ Measurements}
\label{sample}

We use BH mass measurements for 86 AGN from the AGN Black Hole Mass Database\footnote{http://www.astro.gsu.edu/AGNmass/} \citep{Bentz2015}. This database hosts the most complete compilation of RM AGN mass measurements. Most sources include estimates from measuring the lag of the H$\beta$ emission line \citep{Grier2013a}, although estimates from other common lines such as H$\alpha$, C IV, and Mg II are included. Dynamical measurements are also listed, which we use in place of the RM estimate when available.

Table \ref{table1} details the initial calibration sample of 86 AGN. To avoid systematic offsets in mass, we select only AGN with masses estimated from H$\beta$ emission lag measurements (65) \citep[e.g.,][]{Vestergaard2002,Vestergaard2006,Onken2008,Assef2011} or measurements from spatially resolved gas or stellar kinematic measurements (4). Figure \ref{fig:sample_distr} shows the distribution of AGN which make up our final calibration sample. These objects span a wide range of masses from $10^{5.6}$ M$_{\odot}$ to $10^{9.0}$ M$_{\odot}$ and consist mostly of nearby Seyfert galaxies with $z<0.2$. It is important to note that reverberation mapping measurements are heavily skewed towards nearby, bright galaxies, as most RM campaigns have been based on spectra taken with 1-2m telescopes \citep{Bentz2013,Bentz2015}.

\subsection{ASAS-SN Light Curves}
\label{observations}

\begin{figure*}
	\includegraphics[width=\textwidth]{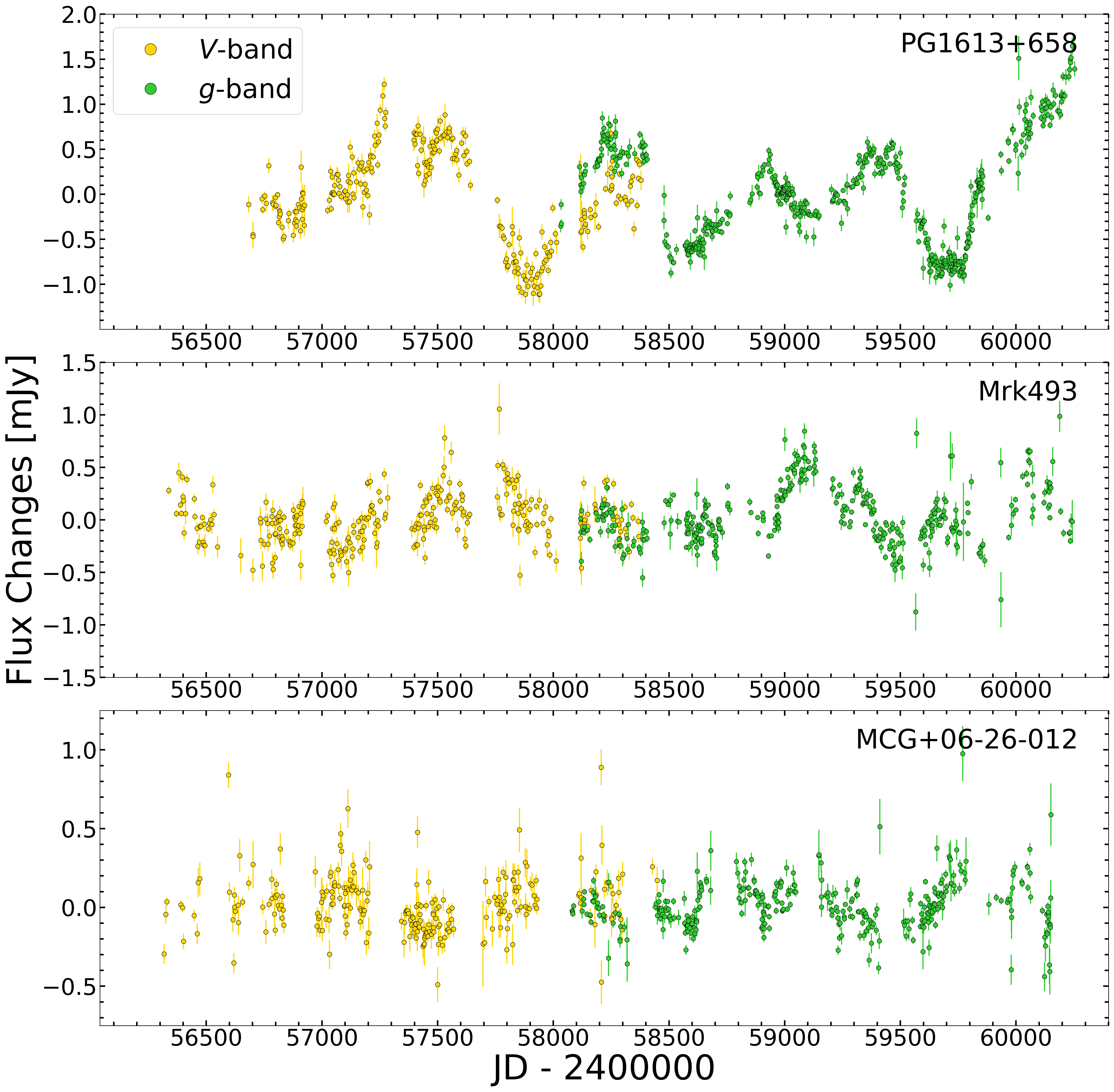}
    \caption{Selected ASAS-SN light curves of target AGN displaying high (top), typical (middle), and low (bottom) variability. Each light curve spans $\sim$11 years and is observed in two optical filters, $V$ (yellow) and $g$ (green). The median flux of each band has been subtracted. Low variability sources, such as the source in the bottom panel, are not included in our final calibration sample (see Section \ref{measure_var}).}
    \label{fig:LCs}
\end{figure*}

\begin{figure*}
	\includegraphics[width=\textwidth]{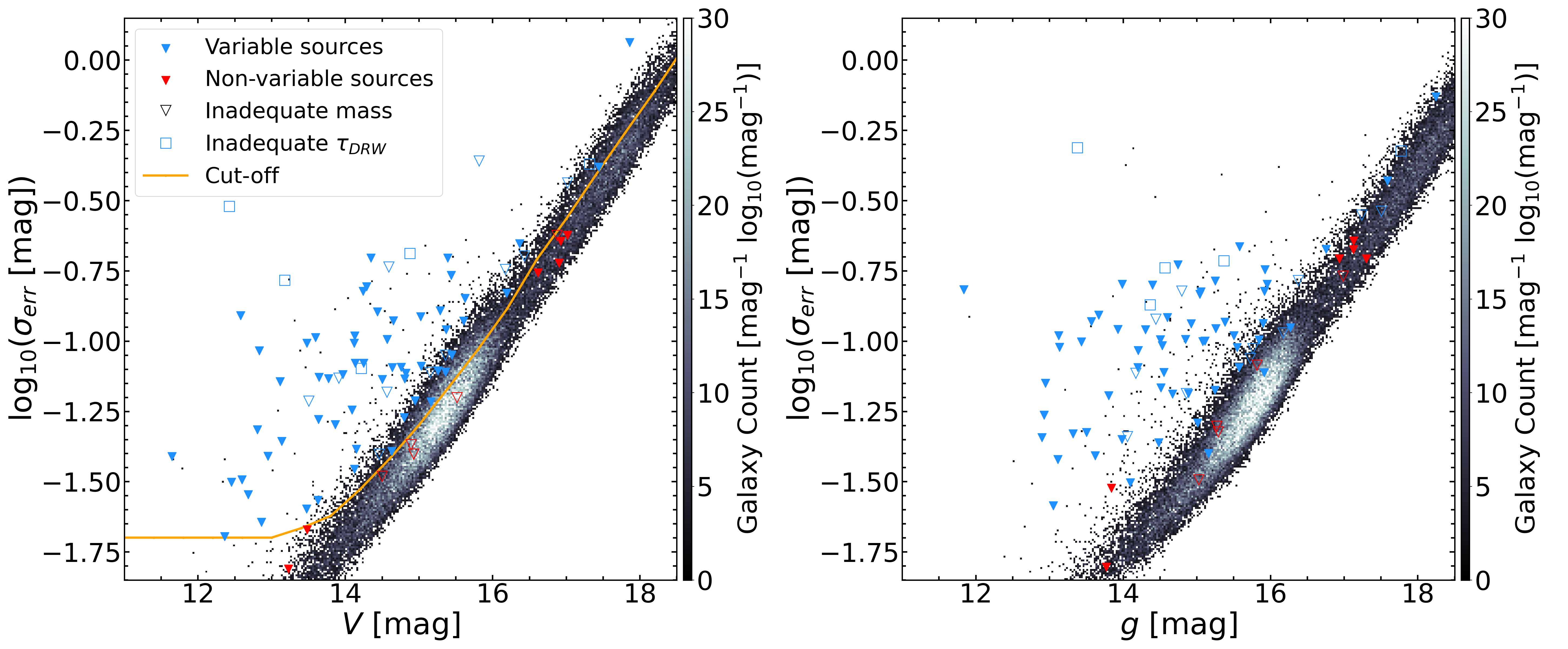}
    \caption{The intrinsic scatter in the calibration AGN light curves as a function of magnitude, compared to the scatter of randomly selected galaxies in $V$-band (left) and $g$-band (right). The orange line shows the 85th percentile light curve scatter for the random galaxies. We show all 86 AGN in our initial calibration sample, where the open triangles are the sources that were removed for not having dynamical or H$\beta$ RM mass measurements (see Section \ref{sample}) and the open squares are removed in Section \ref{est_tau} for poor $\tau \textsubscript{DRW}$ estimates. AGN which are below the variability threshold are shown in red. We only use the light curves of the sources shown as filled blue triangles in the final calibration sample. }
    \label{fig:det_lim}
\end{figure*}

ASAS-SN is a fully-automated, all-sky, transient survey with a typical cadence of $\sim$20 hours in good conditions and a limiting magnitude of $g\sim18.5$ mag. ASAS-SN consists of 20 telescopes across four locations in both the Northern and Southern hemispheres. Each telescope is a Nikon telephoto lens with 14-cm aperture and 4 telescopes share a common mount. Two mounts are located at the Cerro Tololo Inter-American Observatory, and one is located at each of the Haleakalā Observatory, the McDonald Observatory, and the South African Astrophysical Observatory. ASAS-SN observed in the $V$-band filter for the first 5-6 years of survey operations before switching to the $g$-band in 2017/18. The detectors are 2048$^2$ cooled, back-illuminated CCDs with 8\farcs0 pixels and a $\sim$2 pixel FWHM.

During typical ASAS-SN operations, we obtain three dithered 90 second images of each target field and co-add them. We build a reference image from the high quality images and then use the ISIS image subtraction package to subtract the reference image from each new epoch \citep{Alard1998,Alard2000}. A differential light curve is generated by performing aperture photometry on the subtracted images with the \textsc{iraf apphot} package. We add the source flux measured from the reference image back into the subtracted light curve. Lastly, we iteratively sigma-clip to remove points more than 3$\sigma$ away from the median flux of each light curve. 

Figure \ref{fig:sample_distr} shows the distribution of the number of ASAS-SN epochs for the light curves we examine. The typical baseline for an ASAS-SN light curve is 11 years and the median number of epochs is 1,013 for our calibration sources. Figure \ref{fig:LCs} shows full light curves for three of our target AGN. The sources are observed by multiple cameras and there are small calibration offsets between cameras. We describe our handling of camera offsets in Section \ref{est_tau}.  

\subsection{Variability}
\label{measure_var}

To accurately measure $\tau \textsubscript{DRW}$ and $\sigma$ it is important to have a high S/N ratio so that the intrinsic variability is well detected. To quantify the variability, we compare the dispersion in the AGN light curves to that of a large sample of random galaxies. Because the ASAS-SN $V$-band depth is shallower than the $g$-band depth, we use the $V$-band data to conservatively select significantly variable AGN. 

We compare the AGN light curves to the ASAS-SN light curves of over 100,000 galaxies randomly selected from the HyperLEDA catalog \citep{Makarov2014}. We obtained the light curves from the ASAS-SN Sky Patrol V2.0 \citep{Hart2023} database. In Figure \ref{fig:det_lim}, we show the median magnitude and the 1$\sigma$ scatter $\sigma_\text{err}$ of each light curve for both the $V$ and $g$ bands. HyperLEDA contains both AGN and quiescent galaxies, but AGN, the black outliers in the underlying galaxy distribution shown Figure \ref{fig:det_lim}, are significantly outnumbered by quiescent galaxies. Also note that our target AGN were selected from reverberation mapping campaigns and are generally highly-variable. We bin the galaxies into magnitude bins of width 0.4 mag and calculate the 85th percentile of the scatter in each bin to use as a selection cut with linear interpolation between bins. Since there are few galaxies brighter than 13 mag, we extend the dispersion threshold at 13 mag to all brighter galaxies. The 12 sources shown in red are less variable than this minimum variability limit, although some of these sources were already removed from the calibration sample for inadequately measured masses (open triangles). We only use sources shown with filled blue triangles in the final calibration sample. After this variability cut, our calibration sample consists of 63 AGN. An additional 6 sources, shown as open blue squares, are removed in Section \ref{est_tau}.

\subsection{Error Correction}
\label{err_corr}
The maximum likelihood solution for our DRW model is sensitive to the flux uncertainties, where underestimated uncertainties generally lead to underestimates of $\tau \textsubscript{DRW}$. We use a subset of the galaxy light curves from Section \ref{measure_var} to estimate corrections to the pipeline error estimates. \citet{Jayasinghe2018} estimated a correction for this effect for variable stars in ASAS-SN, but here we also account for the angular separation from the camera center to include any effects from vignetting. 

\begin{figure}
    \centering
	\includegraphics[width=\columnwidth]{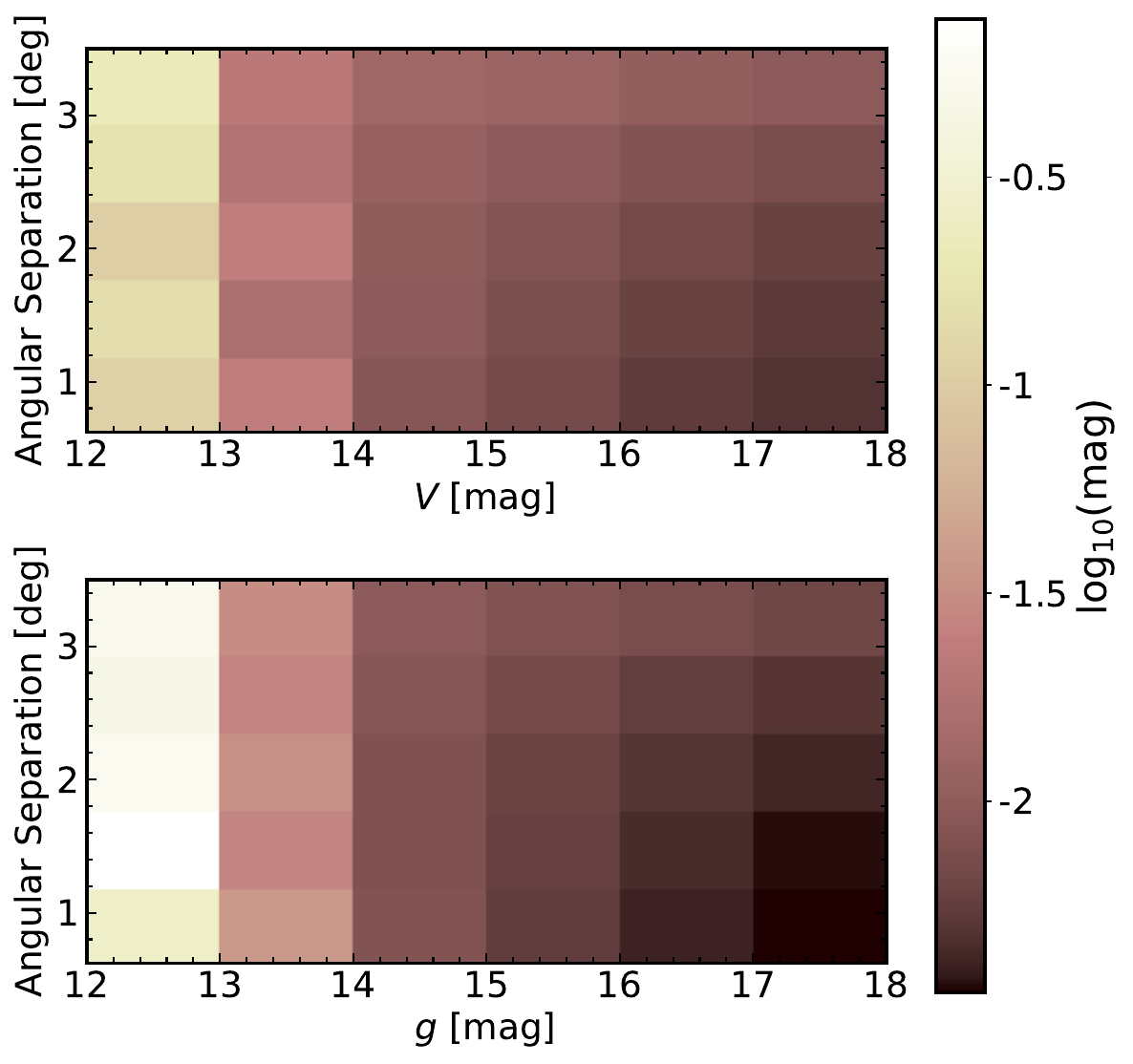}  \caption{Heat maps displaying the error added in quadrature to account for underestimated ASAS-SN uncertainties based on a galaxy's magnitude and angular distance from the ASAS-SN camera pointing center. We calculate these corrections using 50,000+ random galaxy light curves, yielding 88,000+ and 240,000+ correction estimates for $V$ and $g-$band, respectively. We then apply the corrections to each camera's observations before modeling our AGN light curves. Correction values are given in the auxiliary table. }
    \label{fig:error_analysis}
\end{figure}

For each source, we find the extra error that must be added in quadrature to the pipeline errors for a linear fit to have a reduced $\chi^2$ of unity. These are then binned in angular separation from the field center and galaxy magnitude for each camera and filter. We find the median of each bin, as shown in Figure \ref{fig:error_analysis}. In general, the necessary additional error increases with higher angular separation and decreases with magnitude. For each AGN light curve we then add the appropriate correction in quadrature with the reported uncertainties for each camera. Note that at ASAS-SN's resolution ($\sim$16$''$) galaxies are unresolved for all but the nearest, large galaxies, which appear only in the brightest bin. 

\begin{figure*}
    \centering
	\includegraphics[width=\textwidth]{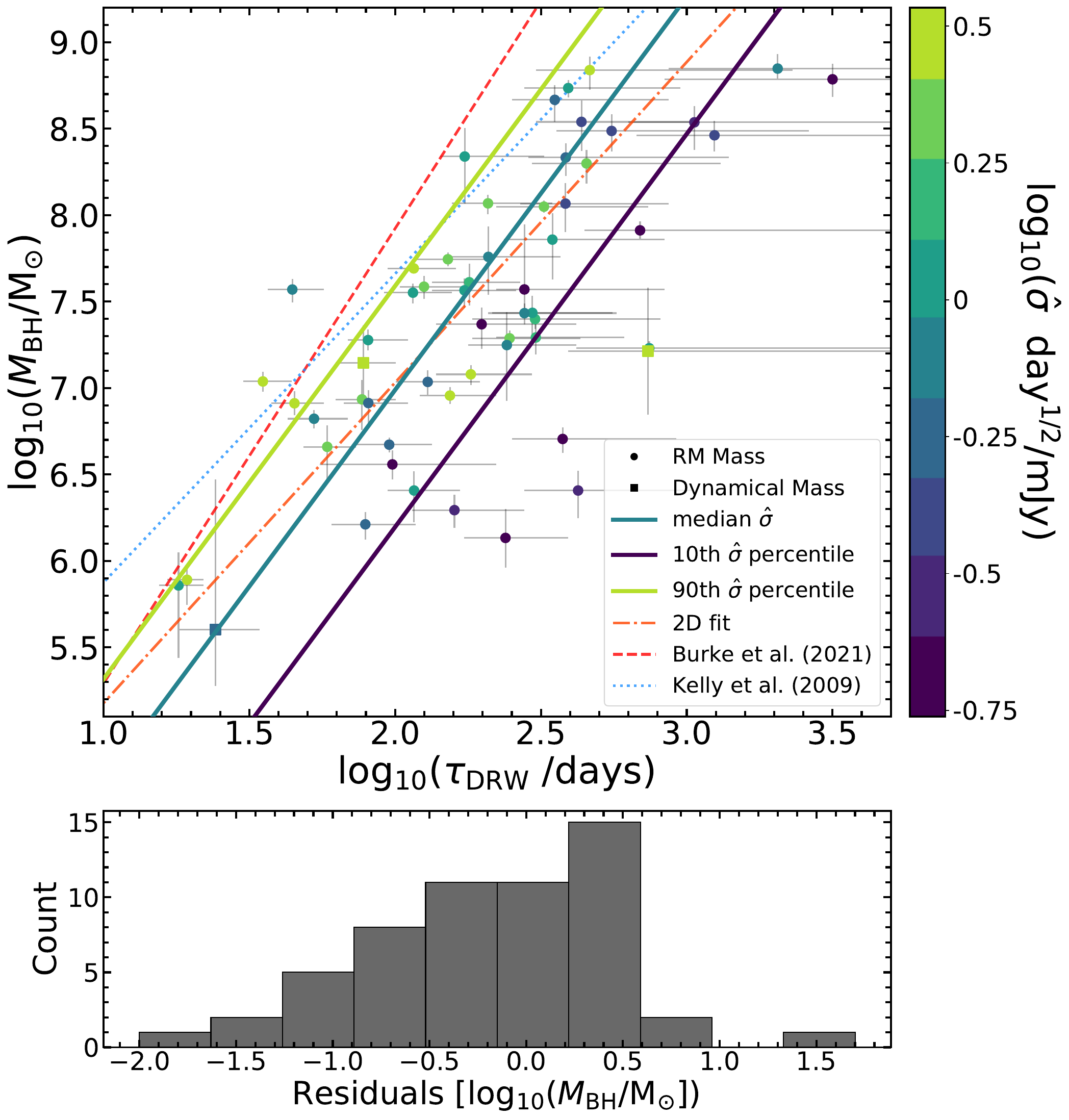}
    \caption{SMBH mass as a function of $\tau \textsubscript{DRW}$ and $\hat{\sigma}$ for the 57 final objects in the calibration sample. The circles are AGN with RM mass measurements and the squares are AGN with dynamical mass measurements from stellar or gas kinematics. We show the best-fit lines for the 2D fit using only $\tau \textsubscript{DRW}$ estimates (orange line) and the multi-variable fit with the 10th, 50th, and 90th percentile values of $\hat{\sigma}$ which are the purple, blue, and green lines respectively corresponding to the color bar. We show the best fit regressions from B21 and KBS09 for comparison, which do not fit the $\hat{\sigma}$ component. We estimate the intrinsic scatter about the regression is 0.39 dex in log$_{10}($M$_{\textrm{BH}} / M_{\odot}$) for the multi-variable fit. In the bottom panel, we show the mass residuals using the median value of $\hat{\sigma}$ which have a dispersion of 0.77 dex.}
    \label{fig:tau_v_mbh}
\end{figure*}

\section{Calibrating the Variability\textendash Mass Relation}
\label{est_tau}

\begin{table*}
  \centering
  \caption{The source name, BH mass, $\tau \textsubscript{DRW}$ estimate, and $\hat{{\sigma}}$ estimate for the AGN in our final sample. All uncertainties are 1$\sigma$, calculated from the log-likelihood distribution. The first 10 sources are shown here, and the full table is included in the electronic edition.}
  \label{table2}
  \renewcommand{\arraystretch}{1.7}
  \begin{tabular}{||c|c|c|c|c|c||}
    \hline
    \textbf{Object} & \textbf{log$_{10}(M_{\textrm{BH}}/\textrm{M}_{\odot})$} & \textbf{log$_{10}(\tau_{\textrm{DRW}}/ \textrm{days})$} & \textbf{log$_{10}(\hat{\sigma}$ /mJy/days$^{1/2}$)} \\
    \hline
    Mrk335 & $7.23^{+0.04}_{-0.04}$  & $2.87^{+2.1}_{-0.26}$  & $0.01^{+0.02}_{-0.02}$  \\
Mrk1501 & $8.07^{+0.12}_{-0.17}$  & $2.58^{+0.32}_{-0.19}$  & $-0.41^{+0.03}_{-0.03}$  \\
PG0026+129 & $8.49^{+0.1}_{-0.12}$  & $2.74^{+0.62}_{-0.25}$  & $-0.45^{+0.04}_{-0.03}$  \\
PG0052+251 & $8.46^{+0.08}_{-0.09}$  & $3.1^{+1.84}_{-0.33}$  & $-0.45^{+0.03}_{-0.03}$  \\
Fairall9 & $8.3^{+0.08}_{-0.12}$  & $2.66^{+0.44}_{-0.21}$  & $0.33^{+0.02}_{-0.01}$  \\
Mrk590 & $7.57^{+0.06}_{-0.07}$  & $1.65^{+0.1}_{-0.1}$  & $-0.13^{+0.03}_{-0.03}$  \\
3C120 & $7.74^{+0.04}_{-0.04}$  & $2.18^{+0.14}_{-0.12}$  & $0.28^{+0.02}_{-0.01}$  \\
Ark120 & $8.07^{+0.05}_{-0.06}$  & $2.32^{+0.21}_{-0.14}$  & $0.28^{+0.02}_{-0.02}$  \\
MCG+08-11-011 & $7.29^{+0.05}_{-0.05}$  & $2.39^{+0.23}_{-0.14}$  & $0.31^{+0.01}_{-0.02}$  \\
Mrk374 & $7.25^{+0.19}_{-0.32}$  & $2.38^{+0.22}_{-0.15}$  & $-0.17^{+0.02}_{-0.03}$  \\
    \hline
  \end{tabular}
\end{table*}

We use a DRW process to estimate the characteristic variability timescale $\tau \textsubscript{DRW}$ and the variability amplitude in flux $\sigma$ for each of our well-measured AGN. Computationally, it is useful to use the variable $\hat{\sigma}^2 = 2 \sigma^2/\tau_{\textrm{DRW}} $ instead of $\sigma$ since there is less covariance between $\tau$ and $\hat{\sigma}$ \citep{Kozlowski2010}. We use the full 11-year light curve consisting of both $V$ and $g-$band flux observations for each AGN with the median flux in each filter subtracted off to calibrate between the two. Following \citet{Kozlowski2010}, the light curves are modeled assuming that the data \textbf{y} consists of a true signal \textbf{s}, a noise component \textbf{n}, and a set of linear parameters \textbf{q} whose effects are defined by a matrix L, such that $\textbf{y} = \textbf{s} + \textbf{n} + L \textbf{q}$. There is one linear parameter for each camera/filter, and they determine the optimal camera calibration offsets. We then find the maximum likelihood solution for the parameters $\tau \textsubscript{DRW}$ and $\hat{\sigma}$ to model this data, solving the system of equations over the entire light curve simultaneously with the linear prediction methods in \citet{Press1992a} and \citet{Rybicki1992}. The uncertainties in the calibration are accounted for in the maximum likelihood solution. This method is computationally inexpensive as the calculations scale linearly with light curve length.

We find the maximum likelihood solution and the 1$\sigma$ parameter uncertainties by evaluating the likelihood surface on a 801$\times$401 grid with $-1\le$ log$_{10}(\hat{\sigma}/ \textrm{mJy/days}^{1/2})\le$ 4 and $-0.5 \le$ log$_{10}(\tau\textsubscript{DRW} / \textrm{days}) \le 5$. Note that $\tau \textsubscript{DRW}$ and $\hat{\sigma}$ are fit parameters from the DRW model and thus do not have a clear physical interpretation; however, they can best be considered respectively as the damping timescale for the light curve to return to its median flux and the rate of change of the variability. All $\tau \textsubscript{DRW}$ estimates are corrected to the AGN rest frame. \citet{MacLeod2010} found that $\tau \textsubscript{DRW}$ and $\sigma$ both depend on the rest wavelength, but given the proximity of peak wavelengths for the $V-$ and $g-$bands, this effect is negligible for the ASAS-SN light curves, as we expect $\tau \textsubscript{DRW}$ to differ by <2\% for most sources and by <5\% for the highest redshift sources in the calibration sample (z$\sim$0.3).

We find reasonable estimates for $\tau \textsubscript{DRW}$ and $\hat{\sigma}$ between $\sim$30\textendash1,000 days and $\sim$0.20\textendash3.5 mJy/days$^{1/2}$, respectively, for most sources. NGC 4151 is the only source with $\hat{\sigma}>40$ mJy/days$^{1/2}$, as its variability amplitude is 2.5 times larger than the median AGN variability amplitude in the calibration sample. A handful of sources approach either $\tau\textsubscript{DRW}\rightarrow0$ or $\tau\textsubscript{DRW}\rightarrow\infty$, and we remove the six sources whose maximum likelihood $\tau \textsubscript{DRW}$ is $<$10 days or $>$10$^{3.5}$ days. With a typical baseline of 11 years, sources with $\tau \textsubscript{DRW} \gtrsim 400$ days may be biased towards shorter values \citep{Kozlowski2021}, but $\tau \textsubscript{DRW}$ estimates longer than this threshold only constitute 20\% of the calibration sample and they have notably larger uncertainties for $\tau \textsubscript{DRW}$. We do not make any cuts on the estimated values of $\hat{\sigma}$ for the calibration sample. Table \ref{table2} lists the values of $\tau \textsubscript{DRW}$ and $\hat{\sigma}$ for the AGN in the final calibration sample of 57 sources.
 
We show the SMBH masses as a function of our estimated values for $\tau \textsubscript{DRW}$ and colored by $\hat{\sigma}$ in Figure \ref{fig:tau_v_mbh}. We confirm previous findings that there is a significant correlation between $\tau \textsubscript{DRW}$ and $M_{\textrm{BH}}$ \citep[e.g., KBS09,][]{MacLeod2010,Burke2020} with a Kendall's tau correlation coefficient of 0.50 and a p-value of 5.0$\times 10^{-8}$. We do not find evidence of a correlation between $\hat{\sigma}$ nor $\sigma$ with $M_{\textrm{BH}}$.

We fit the points in Figure \ref{fig:tau_v_mbh} using the \textsc{LtsFit} package \citep{Cappellari2013} which minimizes the $\chi^2$ statistic while accounting for measurement errors in both independent and dependent variables while allowing for intrinsic scatter within the relation. This analysis assumes symmetric errors, so we take the mean error in each direction for each point. We first fit for the relationship between the SMBH mass and variability timescale only, and find that
\begin{equation}
    \label{eq:2d_fit}
    \textrm{log}_{10}(M_{\textrm{BH}}/ \textrm{M}_{\odot}) = (1.85 \pm 0.20) \times \textrm{log}_{10} \left(  \frac{\tau \textsubscript{DRW}}{200 \textrm{ days}} \right) + 7.59 \pm 0.08 .
\end{equation}
To estimate the intrinsic scatter about the relation, we calculate the amount of error that must be added in quadrature with the uncertainties in the $M_{\textrm{BH}}$ measurements to yield a reduced $\chi^2$ of unity. The $\tau\textsubscript{DRW}$\textendash$M_{\textrm{BH}}$ relation has an intrinsic scatter of 0.44 dex in $\textrm{log}_{10}(M_{\textrm{BH}}/ \textrm{M}_{\odot})$. We investigated correlations between the mass residuals given by this relation and $\hat{\sigma}$, redshift, and Eddington ratio for the calibration sources. We find a significant trend with $\hat{\sigma}$, suggesting that $\hat{\sigma}$ estimates should be included to predict SMBH masses more accurately. More details can be found in Appendix \ref{appendix}. 

Incorporating $\hat{\sigma}$ in the fit yields 
\begin{multline}
    \label{eq:get_mass}
    \textrm{log}_{10}(M_{\textrm{BH}}/ \textrm{M}_{\odot}) = (2.27 \pm 0.20) \times \textrm{log}_{10} \left(  \frac{\tau \textsubscript{DRW}}{200 \textrm{ days}} \right) \\
    + (1.20 \pm 0.20) \times \textrm{log}_{10} \left(  \frac{\hat{\sigma}}{1 \textrm{ mJy/days}^{1/2}} \right) + 7.68 \pm 0.08
\end{multline}
with a modestly reduced scatter of 0.39 dex in $\textrm{log}_{10}(M_{\textrm{BH}}/ \textrm{M}_{\odot})$. We use this relation for the remainder of our analysis.

\begin{figure}
    \centering
	\includegraphics[width=\columnwidth]{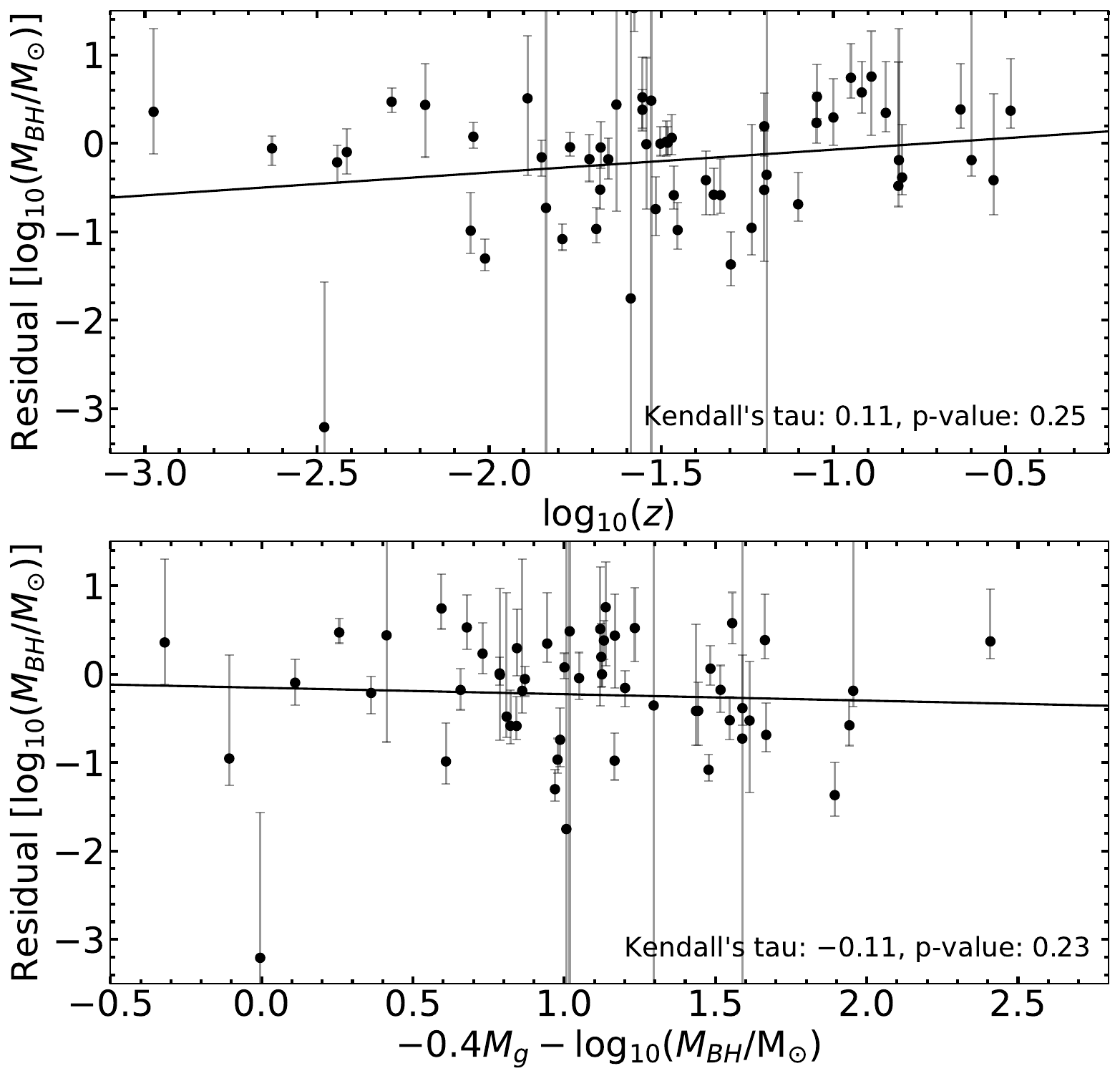}
    \caption{BH mass residuals from Figure \ref{fig:tau_v_mbh} as functions of redshift (top) and a proxy for the Eddington ratio (bottom). We include the Kendall's tau correlation coefficient and corresponding p-value for each set in the bottom right corner of the panels and the black solid line is the least squares regression. Neither set is strongly correlated nor statistically significant. NGC4151 is the strongest outlier given its distinct $\hat{\sigma}$ estimate.}
    \label{fig:res_funcs}
\end{figure}

In Figure \ref{fig:res_funcs}, we show the residuals in log$_{10}(M_{\textrm{BH}}/\textrm{M}_{\odot})$ as a function of redshift and a proxy for the Eddington ratio, $-0.4M_{g}-\textrm{log}_{10}(M_{\textrm{BH}}/\textrm{M}_\odot)$. We include the Kendall's tau correlation coefficient and p-value for each correlation in the bottom right corner of the panels. Neither variable has a statistically significant correlation with the mass residuals from Figure \ref{fig:tau_v_mbh}.

\begin{table*}
  \centering
  \caption{SMBH masses from the field sample calculated with Equation \ref{eq:get_mass}. We list the source name, ASAS-SN ID, right ascension and declination, redshift, apparent $V$-band magnitude, and mass estimate for the first 10 sources. The table is included in its entirety as an ancillary file.}
  \label{table3}
  \renewcommand{\arraystretch}{1.7}
  \begin{tabular}{|c|c|c|c|c|c|c|}
    \hline
    \textbf{Milliquas Source} & \textbf{ASAS-SN ID} &  \textbf{RA (deg)} & \textbf{Dec (deg)} & \textbf{Redshift} & \boldmath{$m_{V}$} & \textbf{log$_{10}$(M$_\textrm{BH}$/M$_{\odot}$)} \\
    \hline
IRAS 21325-6237 & 584115661359 & 324.096 & $-62.400$ & 0.059 & 14.48 & 9.11$^{+4.8}_{-0.63}$\\
PGC 133092 & 498216917200 & 55.861 & $-31.744$ & 0.032 & 14.73 & 5.89$^{+0.76}_{-0.60}$\\
HS 0306+1051 & 489626422401 & 47.236 & $11.054$ & 0.150 & 14.64 & 7.90$^{+1.04}_{-0.30}$\\
MARK 1298 & 446677033375 & 172.320 & $-4.402$ & 0.060 & 14.43 & 8.44$^{+0.96}_{-0.38}$\\
TON S180 & 429497247071 & 14.334 & $-22.382$ & 0.062 & 14.51 & 7.08$^{+0.40}_{-0.19}$\\
KUV 18217+6419 & 393697 & 275.488 & $64.343$ & 0.297 & 13.98 & 8.51$^{+4.97}_{-0.34}$\\
PHL 3953 & 300647927386 & 28.668 & $-27.117$ & 0.151 & 14.90 & 7.73$^{+1.08}_{-0.33}$\\
IC 1198 & 266288556870 & 242.152 & $12.331$ & 0.034 & 14.39 & 7.42$^{+0.63}_{-0.34}$\\
IRAS 13007-1325 & 249109167578 & 195.843 & $-13.693$ & 0.046 & 14.89 & 6.60$^{+0.33}_{-0.21}$\\
S2 0109+22 & 249108589227 & 18.024 & $22.744$ & 0.265 & 14.93 & 6.15$^{+0.36}_{-0.11}$\\
    \hline
  \end{tabular}
\end{table*}

\section{Computing BH Masses For Field AGN}
\label{calc_bh_mass}

Because the variability timescale is well correlated with $M_{\textrm{BH}}$, we use Equation \ref{eq:get_mass} to estimate $M_{\textrm{BH}}$ for field AGN with well-sampled light curves. We retrieved all $V<16$ mag sources from the Million Quasars (Milliquas) catalog \citep{Flesch2021} and obtained their light curves from Sky Patrol V2.0. We exclude known blazars, as their variability is dominated by jet emission and not disk fluctuations (e.g., \citealp{Rees1996,Angel1980,deJaeger23}). We also exclude Type II AGNs where the disk is obscured \citep[e.g.,][]{Antonucci1985,Antonucci1993,Bianchi2012,Netzer2015}. This leaves 17,000 sources above the variability threshold from Section 
\ref{measure_var}.

At ASAS-SN's spatial resolution, crowding by nearby stars can contaminate the AGN light curves and stellar variability can lead to erroneous AGN variability. In particular, giant stars have stochastic variability where both the variability amplitudes and timescales are crudely similar to AGN variability \citep{Treiber23}. To ensure only AGN flux is modeled in our field sample, we use the ATLAS All-Sky Stellar Reference Catalog \citep{refcat} to find and remove sources that have stars within 32$''$ of the AGN's position producing more than 10\% of the flux attributed to the AGN. These criteria leave 1,200 AGN to model with the DRW process.

Following the procedures described in Section \ref{err_corr} and \ref{est_tau}, we find the maximum likelihood $\tau \textsubscript{DRW}$ and $\hat{\sigma}$ solution for each light curve. We again remove sources with rest-frame $\tau\textsubscript{DRW}<$10 days or $\tau\textsubscript{DRW}>$10$^{3.5}$ days. Unlike in the calibration sample, about 20\% of the remaining sources yield $\hat{\sigma} \rightarrow 0$ which we remove from the field sample. For the remaining \NumAGNField{} sources, we estimate the BH masses using Equation \ref{eq:get_mass}. Table \ref{table3} lists these sources and our mass measurements.

\begin{figure}
    \centering
    \includegraphics[width=\columnwidth]{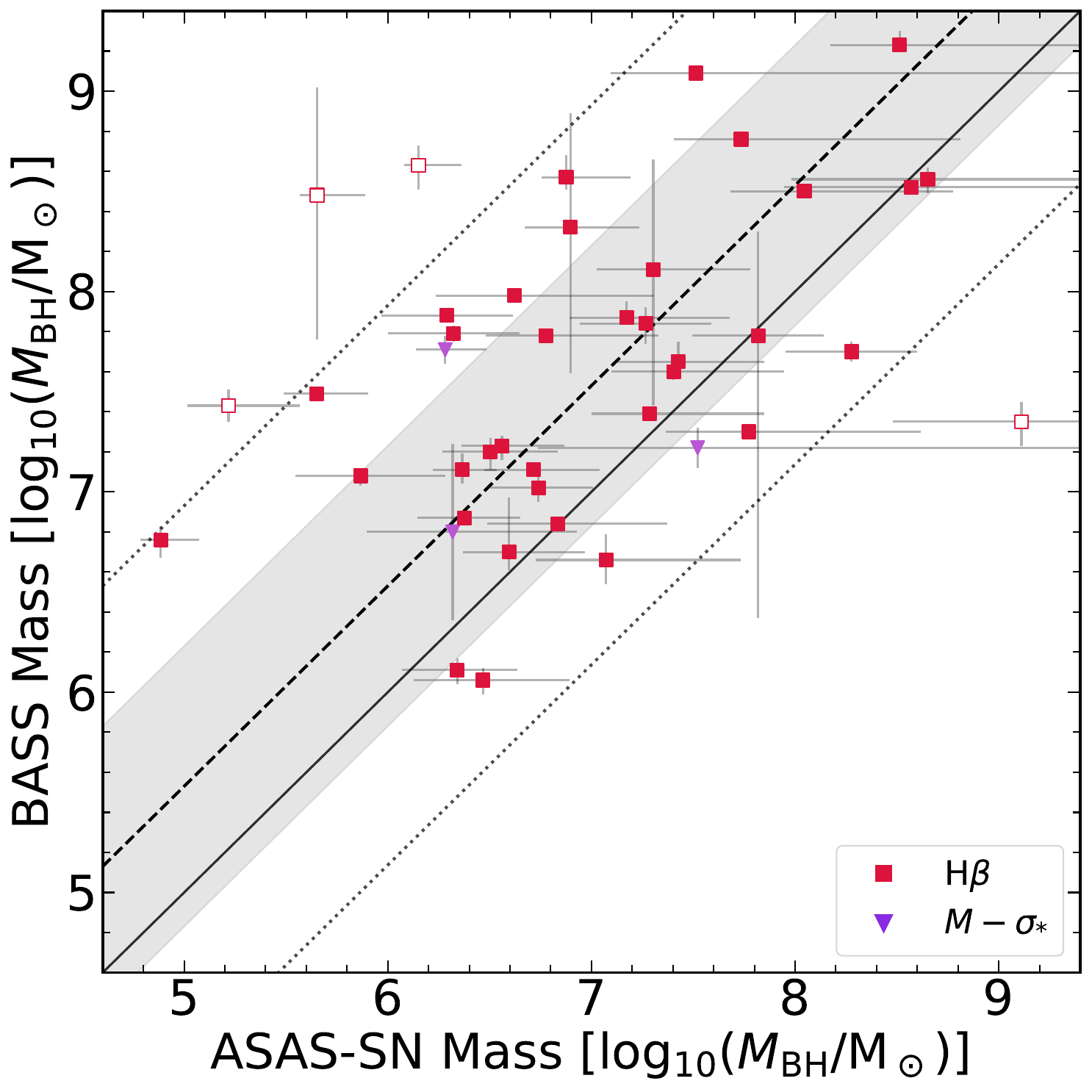}
    \caption{Mass measurements from the BAT AGN Spectroscopic Survey compared to mass estimates from this work. The color indicates the mass measurement method from the BASS survey. The open markers show the 10\% of the data that are 2$\sigma$-clipped, where the dashed lines are the boundaries. The gray line shows the systematic offset between the BASS and ASAS-SN mass estimates and the gray shaded area is the 1$\sigma$ dispersion.}
    \label{fig:massvmass}
\end{figure}

To verify these SMBH mass estimates, we compare them with those that have been measured by the BAT AGN Spectroscopic Survey \citep[BASS;][]{BASSDR2}. BASS has collected over 1,000 optical spectra of AGNs selected from the Swift Burst Alert Telescope \citep[BAT;][]{krimm13} hard X-ray survey. From the optical spectroscopy, the BASS survey has estimated the masses of \OverlapAGN{} AGN in our field sample using single epoch broad-line virial \citep{Mejila2022} and/or stellar velocity dispersion $M_{\textrm{BH}}$\textendash$\sigma_*$ correlation \citep{Koss2022}. When both estimates were available, we used the H$\beta$ broad-line estimate following the priority given in \citet{BASSDR2}. Figure \ref{fig:massvmass} shows the BASS mass against the ASAS-SN masses from Equation \ref{eq:get_mass}. We 2$\sigma$-clip about the line to remove outliers, which are shown with open markers where the boundaries are the dotted lines. The Kendall's tau correlation coefficient is 0.39 with a p-value of $6.0\times10^{-4}$ for the remaining points. We calculate the systematic offset from the BASS measurements, the black dashed line, and 1$\sigma$ dispersion, the gray shaded area. We show a 1:1 line, the black solid line, for comparison. We repeat this calculation using the relations in KBS09 and B21 which are summarized in Table \ref{tab:bass_comp}. The KBS09 relation gives the lowest offset, and the AGN variability plane yields the smallest dispersion.

\begin{table}
    \centering
    \begin{tabular}{l|c|c|c|}
    \hline
         & \textbf{Offset [dex]} & \textbf{Dispersion [dex]}\\
        \hline
        This work & $-$0.53 & 0.70 \\
        KBS09 & 0.34 & 0.87 \\
        B21 & 0.68 & 1.17 \\
    \hline
    \end{tabular}
    \caption{Systematic offset and 1$\sigma$ dispersion measured for the subsample of overlapping sources between BASS and the field AGN sample using the respective variability\textendash mass relations.}
    \label{tab:bass_comp}
\end{table}

\begin{figure}
    \centering
    \includegraphics[width=\columnwidth]{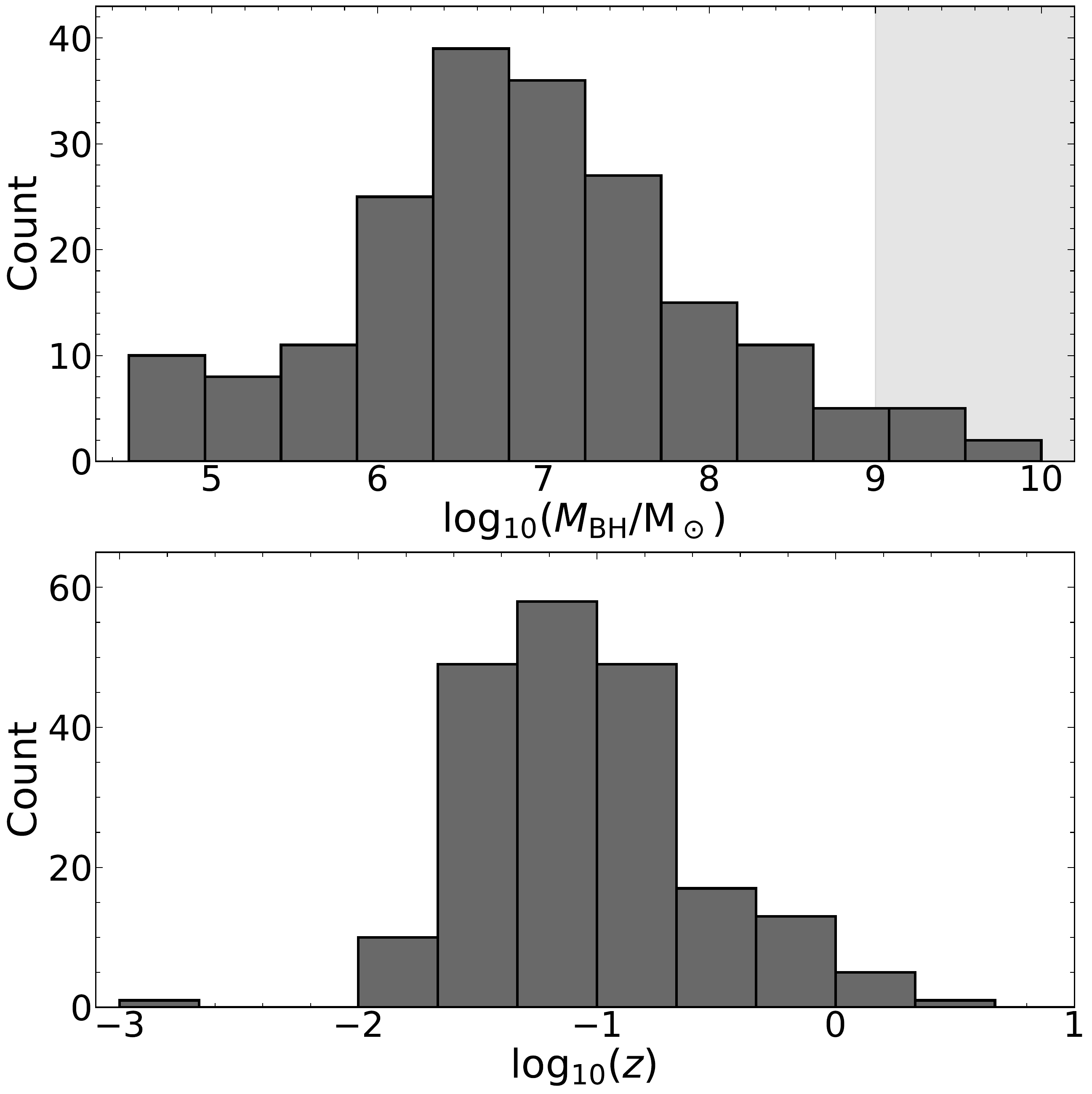}
    \caption{Top: Distribution of mass estimates for \NumAGNField{} field AGN estimated using Equation \ref{eq:get_mass} and presented in Table \ref{table3}. ASAS-SN photometry is able to estimate masses spanning nearly 6 orders of magnitude, although estimates above 10$^{9.0} M_\odot$, the gray shaded region, are more likely to be underestimated given ASAS-SN's baseline. Bottom: Distribution of redshifts for the same field AGN.}
    \label{fig:mass_hist}
\end{figure}

Figure \ref{fig:mass_hist} shows the full mass distribution for the \NumAGNField{} Milliquas sources whose masses we were able to infer with Equation \ref{eq:get_mass} and the corresponding redshift distribution. Our estimates span nearly 6 orders of magnitude and we estimate $M_{\textrm{BH}}$ out to redshift $z=3.9$, although 95\% of the sources are within $z=1$. This sample includes 60 low-mass ($<2 \times 10^6$ M$_{\odot}$) AGN which are significant because low mass sources are frequently too dim to reliably measure with the methods previously discussed. These masses are also likely to be the best constrained, since the corresponding light curves span the most multiples of $\tau \textsubscript{DRW}$ in them \citep{Kozlowski2021}. The current ASAS-SN survey baseline can accurately capture $\tau \textsubscript{DRW}$ corresponding to masses M$\textsubscript{BH}$ < 10$^{9.0}$ assuming Equation \ref{eq:get_mass}. At higher redshifts, accurately estimating $\tau \textsubscript{DRW}$ becomes more difficult, because the rest-frame time baseline diminishes as $z$ increases \citep[e.g.,][]{Kozlowski2021}. 

\begin{figure*}
    \centering
    \includegraphics[width=0.98\textwidth]{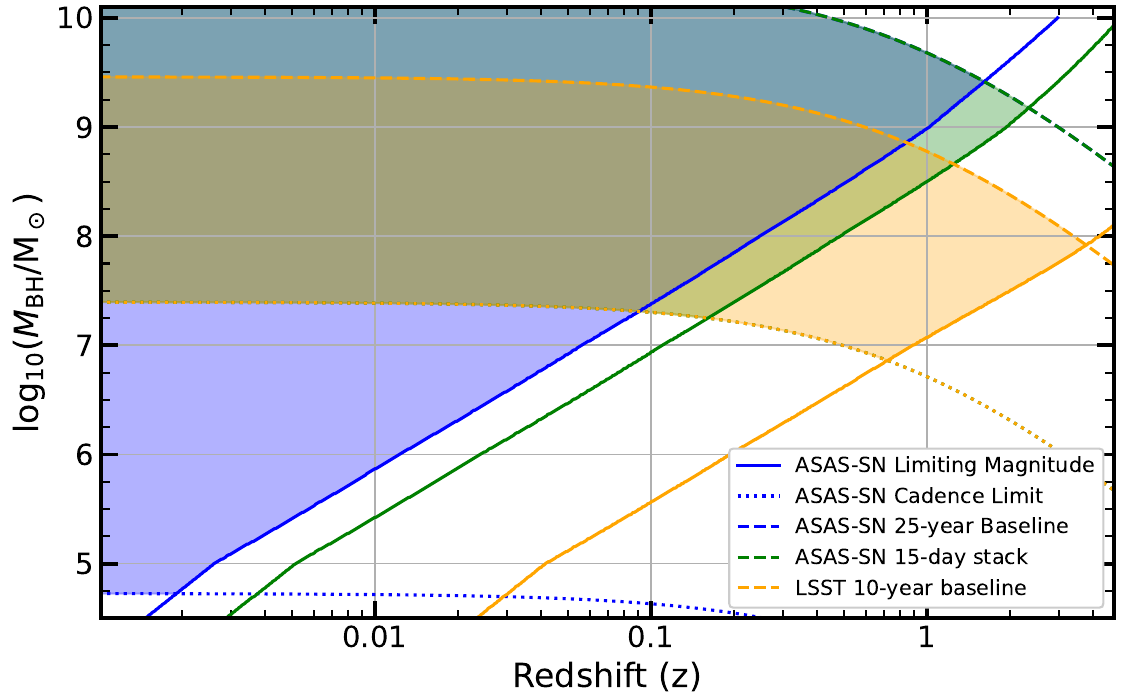}
    \caption{The parameter space in which $\tau \textsubscript{DRW}$ can be measured 12 years into the future with 25 years of ASAS-SN (blue) and 10 years of LSST (orange) $g-$band observations. For the higher $M_\textrm{BH}$ AGN, we also show a 15-day stack of the ASAS-SN survey data in green. The limiting magnitude line (solid) for each survey is taken to be 2 magnitudes above the 5$\sigma$ limits and assumes AGN SEDs generated using the AGNSED model \citep{Kubota18} within xspec v12.14.0h \citep{Arnaud96} as a function of $M_\textrm{BH}$. The SEDs are integrated to estimate the fraction of the bolometric luminosity that will be observed in the $g$-band.  As indicated by the dotted (dashed) lines, the survey cadence (survey length) restricts the minimum (maximum) $M_\textrm{BH}$ able to be measured because of Equation \ref{eq:get_mass}, respectively.  We require that  $\tau\textsubscript{DRW}$ be a factor of 10 higher than the survey cadence and a factor of 10 lower than the survey duration. The shaded regions indicate the SMBHs accessible to each survey at a given redshift.}
    \label{fig:surveycomparison}
\end{figure*}

\section{Discussion}
\label{discussion}

We have demonstrated that measuring optical variability provides reliable $M_{\textrm{BH}}$ estimates with a typical scatter of 0.39 dex for a broad range of SMBH masses. There is a clear relationship between optical variability and $M_{\textrm{BH}}$, which suggests the variability is fueled by physical processes within the accretion disk that correlate with $M_{\textrm{BH}}$. By modeling a homogeneous set of light curves and carefully accounting for the survey systematics, we minimized biases in measuring the correlation between AGN variability and $M_{\textrm{BH}}$, and the scatter of the resulting relation is comparable to that of other scaling relations used to estimate $M_{\textrm{BH}}$.  
 
To our knowledge, this is the first study which fits for both $\tau \textsubscript{DRW}$ and $\hat{\sigma}$ when using AGN variability to estimate SMBH masses. We explored the relationship using both $\sigma$ and $\hat{\sigma}$ with $M_{\textrm{BH}}$ but find a lower dispersion about the plane when fitting for $\hat{\sigma}$. Previous works studying the correlation between AGN variability and $M_{\textrm{BH}}$ focused on the relationship with $\tau \textsubscript{DRW}$ but they did not examine a homogeneous set of light curves. Rather, the light curves and SMBH mass estimates were compiled from across the literature (KBS09; B21). The $\tau \textsubscript{DRW}$ values in KBS09 and B21 were often estimated with light curves taken at different facilities and with distinct baselines, cadences, and/or filters. Here, we analyzed light curves from a single survey to test the correlation on a consistent dataset. The ASAS-SN light curves have both a longer baseline and more epochs than the typical light curve previously used to estimate $\tau \textsubscript{DRW}$, thus allowing us to more accurately sample a wider range of timescale estimates. This dataset also allowed us to account for under-reported uncertainties to correct underestimates of $\tau \textsubscript{DRW}$, since larger uncertainties yield a longer variability timescale as explained in Section \ref{err_corr}. Additionally, both KBS09 and B21 incorporated SMBH mass estimates from common galaxy-SMBH scaling relations, for which the typical scatter is higher than the RM or dynamical measurements adopted in this work. 

As shown in Figure \ref{fig:tau_v_mbh}, we typically measure larger $\tau \textsubscript{DRW}$ values than previous studies. In Section \ref{err_corr}, we found that nearly all sources had underestimated flux uncertainties, suggesting that $\tau \textsubscript{DRW}$ will typically be underestimated without an error correction which likely explains this discrepancy. The lower-mass SMBHs in our sample are also more significantly affected by our error correction procedure (see Figure \ref{fig:error_analysis}) since their intrinsic variability timescale is closest to the ASAS-SN cadence of $\sim$1 day. Low-mass AGNs are less luminous and are therefore more sensitive to photometric uncertainties. 

Other ground-based surveys would likely fail to capture short timescale variability in the noise at timescales close to the survey cadence. The Transiting Exoplanet Survey Satellite \citep[TESS;][]{ricker14} is an exception, with its $\sim 3-30$ minute cadence full-frame images, which are ideal for characterizing the variability of low-mass sources \citep{Burke2020, Treiber23}, although properly accounting for flux uncertainties remains important. For instance, B21 includes the $\tau \textsubscript{DRW}$ estimate of 2.3 days for NGC 4395 using TESS data \citep{Burke2020}. Nevertheless, if our estimate of $\tau \textsubscript{DRW} \sim 24$ days for NGC4395 is correct, a single sector of TESS data would be insufficient to reliably recover this $\tau \textsubscript{DRW}$.

When using AGN variability to estimate $M_{\textrm{BH}}$, the intrinsic scatter about our relation is 0.39 dex which is comparable to other galaxy scaling relations. Several studies have used single-epoch RM mapping, which has a scatter of $\sim$0.6 dex \citep[e.g.,][]{McLure2002,Vestergaard2006}. The correlation between the central $M_{\textrm{BH}}$ and its host galaxy has a scatter of $\sim$0.56 dex for the stellar bulge mass \citep[e.g.,][]{Sahu2019} and $\sim$0.66 dex for its total stellar mass \citep[e.g.,][]{Davis2018}. Thus, the ASAS-SN variability timescale can be used to estimate SMBH masses with better precision than other relations, with the exception of the M\textendash $\sigma_{*}$ relation, which has the tightest correlation with an intrinsic scatter of $\sim$0.3 dex \citep[e.g.,][]{Gultekin2009,Saglia2016,vandenBosch2016}.

Importantly, estimating $\tau \textsubscript{DRW}$ and $\hat{\sigma}$ from public survey light curves avoids the need for costly spectra or host-galaxy decomposition by only requiring photometry. With all-sky surveys like ASAS-SN and well-defined, computationally efficient DRW models, optical variability will become a powerful tool for obtaining $M_{\textrm{BH}}$ estimates. There are other existing long-baseline time-domain surveys like the Asteroid Terrestrial-impact Last Alert System \citep[ATLAS; ][]{Tonry2018}, the Zwicky Transient Facility \citep[ZTF; ][]{Bellm2019}, and the Legacy Survey of Space and Time \citep[LSST, begins soon; ][]{Ivezic2019} on the Vera C. Rubin Observatory (VRO). With readily available photometry, this relation can be easily applied to a very large number of sufficiently bright, well-sampled AGN.

\subsection{Future Variability\textendash$M_{\textrm{BH}}$ Studies}

In approximately 12 years, LSST will provide a 10-year baseline, and ASAS-SN could offer a 25-year baseline. These continuous baselines from single surveys will be critical in minimizing systematic errors in $\tau\textsubscript{DRW}$ measurements, extending the range of probed $M_{\textrm{BH}}$, and enabling studies at higher redshifts. Figure~\ref{fig:surveycomparison} compares the accessible $M_{\textrm{BH}}$ as a function of redshift for both surveys.

For our analysis, we use AGN spectral energy distribtuions (SEDs) as a function of $M_{\textrm{BH}}$ generated using the AGNSED model \citep{Kubota18} within xspec v12.14.0h \citep{Arnaud96}. This model assumes a standard relativistic disk \citep{Novikov73} with three regions: the outer standard disk, the warm Comptonizing region, and the inner hot Comptonizing region. The model incorporates a broad set of parameters to describe these regions, including $M_{\textrm{BH}}$, accretion rate, black hole spin, inclination angle, and the electron temperature, radius, and spectral indices for both the hot and warm Comptonization components. In this study, we assume Eddington-limited accretion with $M_{\textrm{BH}}$ varying from $10^5$ to $10^9$ M$_{\odot}$ in steps of 1 dex. For simplicity, all other parameters are held constant at default values, as outlined in \citet{Kubota18}. We then interpolate these models across $M_{\textrm{BH}}$, apply a standard cosmology to redshift the assumed AGN spectrum, and integrate over the $g$-band to compute apparent magnitudes. We then assume that the light curve variations are detectable by each survey when the computed apparent magnitude is at least 2 magnitudes brighter than the detection limit of each survey ($18.5$ mag for ASAS-SN and $24.5$ mag for LSST). These define the solid lines in Figure~\ref{fig:surveycomparison}.

We further impose survey constraints, requiring the cadence to be $<\tau\textsubscript{DRW}/10$ and the survey duration to exceed $> 10 \tau\textsubscript{DRW}$, using Equation \ref{eq:get_mass} and accounting for time dilation with redshift. We assume a continued daily cadence for ASAS-SN and a 15-day cadence for LSST, representing a $g$-band survey\footnote{Improved cadence may be possible with more complex multi-filter modeling.}. These constraints are reflected as the dotted and dashed lines in Figure~\ref{fig:surveycomparison}. Additionally, since the ASAS-SN cadence is much higher than required for larger $M_{\textrm{BH}}$, we also include the region probed by binning the ASAS-SN data to a 15-day cadence.

It is important to note that Figure~\ref{fig:surveycomparison} is intended to be illustrative. We assume the variability\textendash$M_{\textrm{BH}}$ relation measured in this work does not evolve with redshift, and that it holds for the shorter wavelengths probed by the $g$-band at higher redshifts. \citet{MacLeod2010} found that $\tau \textsubscript{DRW}$ estimates depend on wavelength like $(\lambda_{\textrm{RF}}/4000$ \AA)$^{0.17}$, so shorter rest-frame wavelengths are expected to yield shorter $\tau \textsubscript{DRW}$ estimates. While KBS09 found no correlation between $\tau \textsubscript{DRW}$ and redshift, this trend may not persist at high $z$, where a given filter is probing significantly different accretion disk radii. Additionally, the assumptions of Eddington luminosity and no extinction clearly does not probe the average SMBH. Nevertheless, Figure \ref{fig:surveycomparison} demonstrates that light curve variability is a promising tool for identifying and characterizing the high- and low-mass SMBHs extremes in the local universe with long-term, fast cadence surveys like ASAS-SN.  Figure \ref{fig:surveycomparison} also shows that the combination of 25-year ASAS-SN and the 10-year LSST survey on the VRO will probe $\textrm{log}_{10}({M_\text{BH}/\textrm{M}_\odot}) > 7$ at $z \sim 1$. And finally, the depth of the LSST survey on the VRO will enable it to probe $M_\text{BH} \sim 8.0$ out to $z \sim 4$.

\section{Summary and Conclusions}
\label{conclusion}

We examined 57 homogeneously obtained ASAS-SN light curves spanning $\sim$11 years with a typical cadence of $\sim$1 day for AGN with well-measured SMBH masses to calibrate the correlation between AGN optical variability with $M_{\textrm{BH}}$. We modeled the light curves using a DRW process and combined the variability timescale and amplitude estimates with $M_{\textrm{BH}}$ measurements from reverberation mapping and dynamical measurements to minimize systematic biases. We measured a relationship with an intrinsic scatter of 0.39 dex, which is comparable to or better than other scaling relations for SMBH masses. We then applied this relation to a field sample from the Milliquas catalog and compared our estimates to broad-line and velocity dispersion mass measurements from the BASS survey for the sub-sample of sources in both catalogs.

We generated mass estimates for \NumAGNField{} field AGN, whose distribution is shown in Figure \ref{fig:mass_hist}. This application of the variability\textendash $M_{\textrm{BH}}$ relation is to demonstrate its function, and further analysis of the field sample is out of the scope of this paper. Of the inferred AGN masses, 60 are $<2\times10^6$ M$_{\odot}$. These are particularly interesting because low-mass AGN are difficult to study since they are less luminous, making high-quality observations for other mass measurement methods like RM difficult to obtain. ASAS-SN could increase the observational cadence for these low-mass AGN. A higher sampling cadence of 2 \textendash{} 3 times per day is feasible, and sampling at the increased cadence for $\sim$1 year would lead to substantially improved $\tau \textsubscript{DRW}$ estimate for $M_{\textrm{BH}}\sim10^5-10^6$M$_{\odot}$ AGN. 

Future mass calibration efforts need more RM or dynamical measurements, especially to better sample lower mass AGN. Our calibration sample contained only 3 low-mass sources. NGC 4395 has been well-studied with RM, stellar dynamics, and gas dynamics, and while these measurements agree within their uncertainties, they span an order of magnitude in BH mass.

Using the variability\textendash$M_{\textrm{BH}}$ relation only requires photometric data which is increasingly available from the expanding number of all-sky surveys. In $\sim$12 years, the upcoming LSST survey will complete its 10-year 6-band survey and the ASAS-SN survey will have a 25-year baseline. As shown in Figure~\ref{fig:surveycomparison}, LSST will characterize AGN variability out to $z\sim3$ for AGN with $M_{\textrm{BH}} \sim 10^{8.0}$ M$_{\odot}$. With a 25-year baseline ASAS-SN will be able to recover variability timescales on the order of 2.5 years, which for the calibration in Equation \ref{eq:get_mass} corresponds to $M_{\textrm{BH}} \approx 10^{9.9}$ M$_{\odot}$. Finally a higher-cadence ASAS-SN sub-survey could probe down to $M_{\textrm{BH}} \sim 10^5$ M$_{\odot}$, enabling ASAS-SN to reliably probe nearly 5 dex in mass.


\section*{Acknowledgements}
AT acknowledges support from Research Experience for Undergraduate program at the Institute for Astronomy, University of Hawai'i-Mānoa funded through NSF grant \#2050710. AT would like to thank the Institute for Astronomy for their hospitality during the course of this project.

JTH is supported by NASA grant 80NSSC23K1431. The Shappee group at the University of Hawai'i is supported with funds from NSF (grants AST-1908952, AST-1911074, \& AST-1920392) and NASA (grants HST-GO-17087, 80NSSC24K0521, 80NSSC24K0490, 80NSSC24K0508, 80NSSC23K0058, \& 80NSSC23K1431). CSK is supported by NSF grants AST-1907570, 2307385, and 2407206. 

We thank the Las Cumbres Observatory and their staff for its continuing support of the ASAS-SN project. ASAS-SN is funded  by Gordon and Betty Moore Foundation grants GBMF5490 and GBMF10501, and the Alfred P. Sloan Foundation grant G-2021-14192. 

Development of ASAS-SN has been supported by NSF grant AST-0908816, the Mt. Cuba Astronomical Foundation, the Center for Cosmology and AstroParticle Physics at the Ohio State University, the Chinese Academy of Sciences South America Center for Astronomy (CAS- SACA), the Villum Foundation, and George Skestos. TAT is supported in part by Scialog Scholar grant 24216 from the Research Corporation. Support for JLP is provided in part by FONDECYT through the grant 1151445 and by the Ministry of Economy, Development, and Tourism’s Millennium Science Initiative through grant IC120009, awarded to The Millennium Institute of Astrophysics, MAS. 

HT acknowledges support by the National Science Foundation Graduate Research Fellowship Program under Grant DGE-2039656. Any opinions, findings, and conclusions or recommendations expressed in this material are those of the authors and do not necessarily reflect the views of the National Science Foundation.

Parts of this research were supported by the Australian Research Council Discovery Early Career Researcher Award (DECRA) through project number DE230101069.

This work is based on observations made by the ASAS-SN telescopes. The authors wish to recognize and acknowledge the very significant cultural role and reverence that the summit of Haleakalā has always had within the indigenous Hawaiian community.  We are most fortunate to have the opportunity to conduct observations from this mountain.

\section*{Data Availability}
The datasets were derived from sources in the public domain: ASAS-SN Sky Patrol Photometry Database (http://asas-sn.ifa.hawaii.edu/skypatrol/), AGN Black Hole Mass Database (http://www.astro.gsu.edu/AGNmass/).


\bibliographystyle{mnras}
\typeout{}
\bibliography{main}


\appendix

\section{2D Fit Analysis}
\label{appendix}

\begin{figure}
    \centering
    \includegraphics[width=0.8\linewidth]{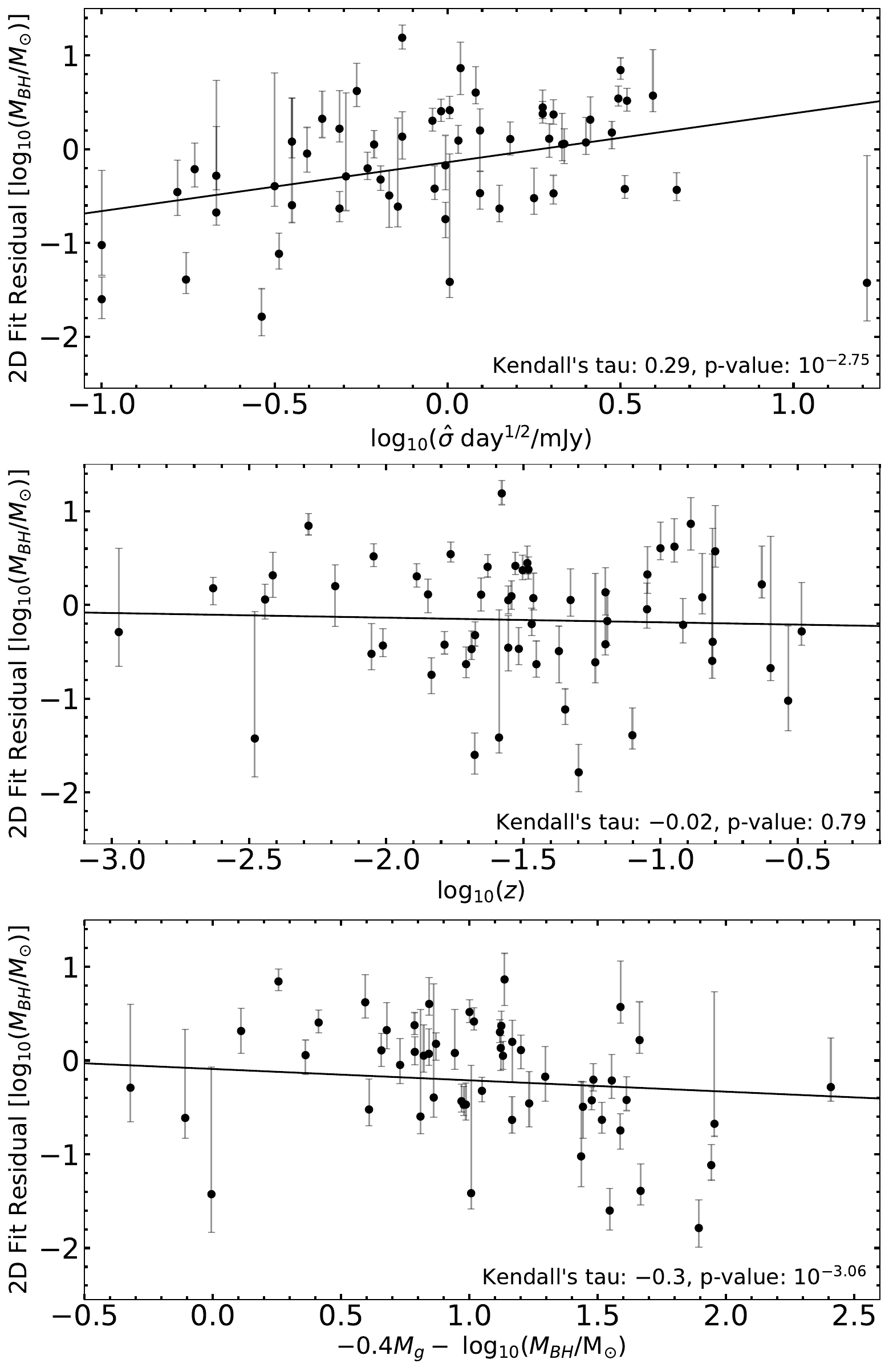}
    \caption{SMBH mass residuals given by Equation \ref{eq:2d_fit} as a function of $\hat{\sigma}$ (top), redshift (middle), and Eddington ratio proxy (bottom). There is a significant correlation between the residuals and $\hat{\sigma}$, motivating the inclusion of variability amplitude in the final AGN variability plane fit (Equation \ref{eq:get_mass}). There is also a significant correlation between residuals and Eddington ratio for the 2D fit, but this disappears in the final fit after incorporating $\hat{\sigma}$ estimates.}
    \label{fig:2d_M_res}
\end{figure}

Previous studies have examined the relationship between AGN variability and SMBH mass focusing on the variability timescale and amplitude independently (KBS09, B21). We first fit for the relationship between $\tau \textsubscript{DRW}$ and SMBH mass in Equation \ref{eq:2d_fit} and find an intrinsic scatter in SMBH mass of 0.44 dex. We calculate the mass residuals given by this equation and the RM or dynamical masses as a function of $\hat{\sigma}$, redshift, and Eddington ratio proxy $-0.4 M_g-\text{log}_{10}(M_{\text{BH}}/\text{M}_\odot)$, shown in Figure \ref{fig:2d_M_res}. We find a significant correlation between mass residuals and $\hat{\sigma}$, shown in the top panel, where shorter $\hat{\sigma}$ estimates correlate with more underestimated SMBH masses. We correct for this by incorporating $\hat{\sigma}$ estimates into the fit for SMBH mass, given by Equation \ref{eq:get_mass}, which has a 10\% reduced scatter compared to the fit without $\hat{\sigma}$ estimates. We also find a significant correlation between mass residuals and Eddington ratio when using the Equation \ref{eq:2d_fit}, which might suggest that systems with higher accretion rates will tend to yield overestimated masses measured by optical variability. However, this correlation disappears when the variability amplitude is used in the fit (see Figure \ref{fig:res_funcs}). There is no correlation between mass residuals and redshift. 

\bsp	
\label{lastpage}
\end{document}